\title{\LARGE \textbf{Towards small and accurate convolutional neural networks for acoustic biodiversity monitoring}}  

\date{\today} %

\documentclass[12pt]{article}

\usepackage[
	sorting = none ,
	backend = biber,
	style = numeric, 
	]{biblatex}
\usepackage{graphicx}
\usepackage[ 
	colorlinks=true,
	pdfborder={0 0 0},
	linkcolor=black
	]{hyperref}
\usepackage{csvsimple}
\usepackage{authblk}
\usepackage{float}
\usepackage{xcolor}
\usepackage{lineno}
\usepackage{subcaption} 
\usepackage[singlelinecheck=false]{caption}
\usepackage[a4paper,
	bindingoffset=0.0in,
	left=1in,
	right=1in,
	top=1in,
	bottom=1in,
	footskip=.25in]{geometry}

\author[1]{Serge Zaugg}
\author[1]{Mike van der Schaar}
\author[1]{Florence Erbs}
\author[1]{Antonio Sanchez}
\author[1]{Joan V. Castell}
\author[2]{Emiliano Ramallo}
\author[1*]{Michel André}

\affil[1]{Laboratory of Applied Bioacoustics - Universitat Politècnica de Catalunya, Barcelona, Spain}
\affil[2]{Instituto de Desenvolvimento Sustentável Mamirauá, Tefé, AM, Brazil}
\affil[*]{Corresponding author: michel.andre@upc.edu}

\graphicspath{	{figs/} {figs_dynamic/} }

\begin{document}
	\maketitle
	
	\begin{abstract}
		
		Automated classification of animal sounds is a prerequisite for large-scale monitoring of biodiversity. Convolutional Neural Networks (CNNs) are among the most promising algorithms but they are slow, often achieve poor classification in the field and typically require large training data sets. Our objective was to design CNNs that are fast at inference time and achieve good classification performance while learning from moderate-sized data. Recordings from a rainforest ecosystem were used. Start and end-point of sounds from 20 bird species were manually annotated. Spectrograms from 10 second segments were used as CNN input. We designed simple CNNs with a frequency unwrapping layer (SIMP-FU models) such that any output unit was connected to all spectrogram frequencies but only to a sub-region of time, the  Receptive Field (RF). Our models allowed experimentation with different RF durations. Models either used the time-indexed labels that encode start and end-point of sounds or simpler segment-level labels. Models learning from time-indexed labels performed considerably better than their segment-level counterparts. Best classification performances was achieved for models with intermediate RF duration of 1.5 seconds. The best SIMP-FU models achieved AUCs over 0.95 in 18 of 20 classes on the test set. On compact low-cost hardware the best SIMP-FU models evaluated up to seven times faster than real-time data acquisition. RF duration was a major driver of classification performance. The optimum of 1.5 s was in the same range as the duration of the sounds. Our models achieved good classification performance while learning from moderate-sized training data. This is explained by the usage of time-indexed labels during training and adequately sized RF. Results confirm the feasibility of deploying small CNNs with good classification performance on compact low-cost devices.
	
	\end{abstract}

	\newpage

	\tableofcontents
	
	\newpage
	
	\section{Keywords}	
	Bio-acoustics, Biodiversity, Ecological monitoring, Machine learning, Deep learning, Sensors
	
	\section{Introduction}

	Animal species from diverse groups such as birds, mammals, frogs, fishes, and insects produce sounds. The automated and continuous detection of multiple species at multiple sites would be a valuable asset for the monitoring of large-scale changes in biodiversity \cite{gibb2019emerging}. The present work is part of a larger project (www.projectprovidence.org) which proposes to implement long-term and large-scale acoustic monitoring of Amazon ecosystems. At project start, labeled reference data for training and testing of supervised classification processes was not available, which is likely to be the case for many bioacoustic projects. Considering the high burden imposed by labeling, we proposed to use moderate-sized training data.
	
	Acoustic monitoring in remote areas like the Amazon forest is challenging. The periodic manual retrieval of data storage devices is not practical. Transmission of data over wireless channels is a promising solution but due to the limited bandwidth this is only feasible if transmitted data volume is heavily reduced. An efficient way to achieve data reduction is to transmit only predictions obtained locally on devices with limited computational capacity. To address this, we evaluated multiple Convolutional Neural Networks (CNNs) to find models that are both fast at inference time and achieve good predictive performance.  
	
	CNNs require considerable quantities of labeled data, both for testing and training. Accurate labeling of the test data is critical for a good estimate of predictive performance while accurate labeling of training data generally leads to better models. However, for many species no labeled acoustic data exists. In many monitoring projects, data will have to be labeled from scratch. The ability to work with moderate-sized datasets would be a major asset for such studies.
	
\begin{figure}[H]
		\fbox{\includegraphics[width=0.9\textwidth]{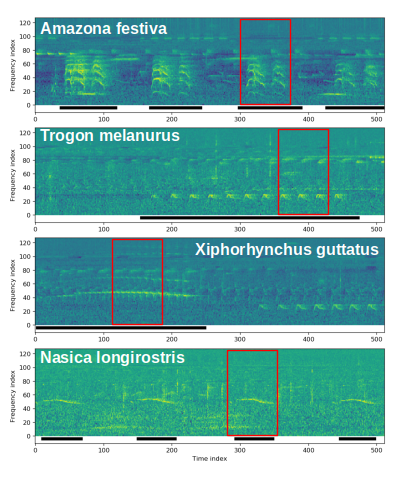}}
		\caption{Spectrograms used as CNN input. Four sound-types shown here; all 20 in Supplement. Time-indexed labels used as target during training shown as black bars at the bottom of each spectrogram. Manually selected Regions of 1.5 seconds shown as red boxes to allow visual comparison with the time scale of acoustic patterns.}
		\label{fig:spectrosmain}
	\end{figure}

	The labels used in this study mark the start and end of a target sound and are encoded such that this information can be used during network training (See Figure \ref{fig:spectrosmain}). They are called time-indexed labels hereafter. CNNs are prone to "Shortcut learning" and consequently they tend to perform poorly on independent test data \cite{geirhos2020shortcut}. In our setting "Shortcut learning" could occur if patterns from sounds in the close vicinity of the target are used for learning instead of the target sound itself. This risk is reduced by explicitly defining the start and stop time of the targets.

	Compared to photograph images, spectrogram patterns are less complex (blurry, one channel, fixed time scale, fixed frequency location). A key assumption of this study was that CNNs with only a few convolutional layers have sufficient expressive power to capture these patterns. In acoustic monitoring the difficulty often arises from the diversity of background noise and presence of other interfering sounds. Simpler CNNs tend to be less prone to over-fitting and under-specification \cite{damour2020underspecification}. Over-fitting is especially critical when small training sets have to be used and under-specification could be critical in the presence of correlated interfering sounds. Simpler models are expected to be more robust when deployed in the real-world. 
	
	Until recently, research on bioacoustic signal classification has focused on developing feature extraction procedures specifically for a particular problem followed by conventional supervised classification methods. Reviews are available in \cite{bittle2013review, malfante2018automatic, priyadarshani2018automated, xie2018frog, gibb2019emerging, stowell2021computational}. Bioacoustic signals are very variable depending on the targeted animals and environments. Generally, feature extraction procedures developed for a particular problem cannot be transferred to other problems. In recent years, Deep Neural Networks (DNNs) have been used successfully for bioacoustic classification tasks, mostly for birds but also other animal groups \cite{lebien2020pipeline, }. DNNs are generally considered superior in terms of classification performance \cite{shiu2020deep} and do not require the tedious development of feature extraction procedure. The majority of bioacoustic studies used CNNs with 2D convolutional filters \cite{grill2017two, kahl2017large, salamon2017fusing, sevilla2017audio, incze2018bird, lasseck2018audio, liu2018classification, mac2018bat}  and a few used recurrent neural networks \cite{himawan2018deep, ibrahim2018automatic}. In most studies, the acoustic signal was transformed to 2D arrays via the Short Time Fourier Transform (STFT). The log-transformed frequency dimension (Mel-spectrogram) was typically used as input to the DNNs. While many studies used large imaging CNNs, some studies achieved good performance with smaller CNNs consisting of 2-9 convolutional layers \cite{grill2017two,  kahl2017large,  liu2018classification,  mac2018bat}.
	
	\section{Data and labeling}
	
	\subsection{Data sources}
	
	Acoustic data from the Mamirauá Sustainable Development Reserve (MSDR) in Amazonas, Brazil were recorded with Wildlife SM4 recorders in two periods and stored as one-channel wave files. During Period 1 (July 2016 to April 2017) data were recorded continuously (30 min wave file every 30 min) in 4 sessions of between 3 and 9 days. One minute was randomly extracted from each 30 min file to be manually annotated. This data was used as test set. During Period 2 (August to October 2017) data were recorded with a regular duty cycle (5 min wave files every hour). One minute was randomly extracted from each 5 min file to be manually annotated. This data was used as training set. The randomly extracted data from both periods cover all hours of day and night and they are representative of soundscapes expected in the MSDR. Due to very low counts of some target sounds in the training set, additional data of the 20 bird species were downloaded from the xeno-canto repository (www.xeno-canto.org, download criteria see Supplement). These are recordings performed or selected by humans, sometimes with directional microphones. The sounds typically exhibit a high SNR compared to continuous recordings from the wild. They were included in the training data. 
	
	\subsection{Data labeling}
	
\begin{figure}[H]
		\fbox{\includegraphics[width=0.9\textwidth]{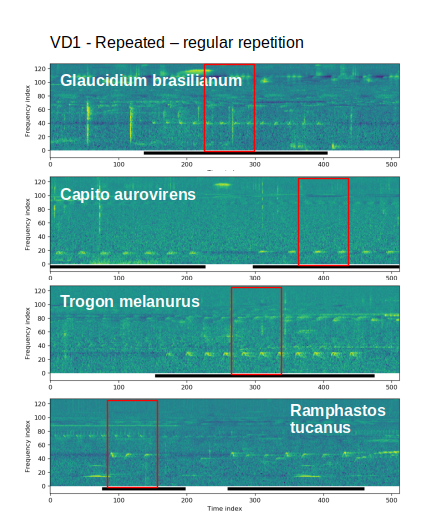}}
		\caption{Spectrograms used as CNN input. Sound-types with regular repetition pattern}
		\label{fig:spe04}
	\end{figure}

\begin{figure}[H]
		\fbox{\includegraphics[width=0.9\textwidth]{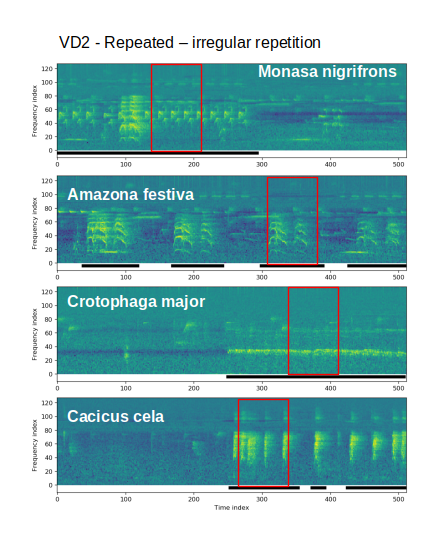}}
		\caption{Spectrograms used as CNN input. Sound-types with non-regular repetition pattern}		
		\label{fig:spe05}
	\end{figure}

	A sound-type is defined here as an animal vocalization that can be consistently recognized by human experts and characterized with a few examples. Most bird species produce multiple sound-types which are acoustically distinctive and have a specific behavioral function (e.g. territorial song, contact call, alarm call). Sound-types from 20 bird species that were identified as relevant for bio-acoustic monitoring in the MSDR were searched and labeled for this study (Table 1 and Appendix for details). The start/end point of each sound were marked with bounding boxes with a dedicated labeling tool that allowed in-depth visual and acoustic scrutiny of the sounds. The boxes also included lower and upper frequency bounds but they were not used in this study. The 20 sound-types were assigned to 5 groups according to their typical duration. Twelve sound-types had a fixed duration (Groups FD1, FD2, FD3) and 8 had a variable duration (Groups VD1 and VD2). The sound-types are summarized in Table 1 and illustrated in Figure \ref{fig:spectrosmain}, and in supplementary Figures \ref{fig:spe01}, \ref{fig:spe02}, \ref{fig:spe03}, \ref{fig:spe04}, \ref{fig:spe05}.
	
	Test data (period 1) were exhaustively searched for the 20 sound-types with the labeling tool. Each single segment was scrutinized even if it contained only background noise and all signals that could be discerned were labeled. Training data (Period 2 and xeno-canto) were labeled with computer aided labeling. Candidate regions where first identified via model-based predictions with a low decision threshold to minimize the number of false negatives. Candidate regions were then manually labeled by a human expert with the dedicated tool.    
	
	\subsection{Data pre-processing}
	
	Acoustic data from MSDR were recorded as waveforms (WAV format) with a sampling rate of 48000 samples per second (sps) and data from xeno-canto were downloaded as MP3 and converted to WAV with a sampling rate of 48000 sps. All files were processed by segments of 10 seconds. For the training set this resulted in 10242 segments from Period 2 plus 1099 segments from xeno-canto. For the test set this resulted in 5635 segments from Period 1 (detail in Appendix Table 7). The high sampling rate was motivated by the detection of ultrasonic sounds of some mammals which are not the focus of this study. All bird sounds used here were located in frequencies below 5000 Hz. Frequencies below 100 Hz were dominated by ambient noise (wind, water). Therefore, the operational range of the classifier was set to 100-5000 Hz. The waveforms were pre-processed by segments of 10 seconds in several steps to obtain standardized Mel-spectrograms: (1) compute the Short Time Fourier Transform (STFT, Hann window of 2048 samples = 0.04 s, 54\% overlap), (2) select  frequency bins between 100 Hz and 5000 Hz, (3) apply absolute value and log to the STFT, (4) equalize STFT by subtracting the per-frequency median, (5) apply 128 Mel-scaled filters to the frequency dimension of the STFT. Mel-scaling emphasizes the resolution of lower relative to higher frequencies. This allows a considerable reduction in array size while conserving information content of the spectrograms. (6) Finally, normalize amplitude to have zero mean and unit standard deviation. The result is a matrix of 512 time bins by 128 frequency bins. 
	
	The original annotations were the start/end point of each of the 20 sound-types and as such they cannot be directly used to train CNNs. The annotations were transformed into binary matrices of size 512 by 20 (time by number of classes) that encode the presence of each sound at a particular time. If necessary the time dimension of these matrices was down-sampled via block-wise max pooling to match the size of the CNN's output. These binary matrices will be referred to as time-indexed labels hereafter. Visual representations of spectrograms and time-indexed labels are shown in Figure \ref{fig:spectrosmain} and in supplementary Figures \ref{fig:spe01}, \ref{fig:spe02}, \ref{fig:spe03}, \ref{fig:spe04}, \ref{fig:spe05}. 
		
	\section{Methods}
	
	\subsection{Receptive field}
	
	The Receptive Field (RF) of a CNN is a region of the input array (in our case a spectrogram) that is connected to one unit of the output array. In other words, how much of the input spectrogram can one output unit 'see'. Intuitively, if it sees too long regions it might wrongly learn from surrounding background signals. If it sees too short regions, it will not leverage the full pattern of a sound during the learning process. In this article the focus is on the time dimension where the input has 512 units (10 sec) and the output between 1 and 512 units, depending on the model. For our models we report the Maximum Receptive Field (MRF, Table 4) which can be derived by arithmetic calculation (see for example eq. 2 in \cite{araujo2019computing}). Recent research has defined the Effective Receptive Field and shown that units closer to the center of the receptive field have a larger impact on the output value \cite{luo2016understanding}. Therefore the MRF that we report here should be considered an upper bound on Effective Receptive Field size. 
	
	\subsection{Simple Frequency Unwrapping CNNs}
	
\begin{figure}[H]
		\fbox{\includegraphics[width=1.0\textwidth]{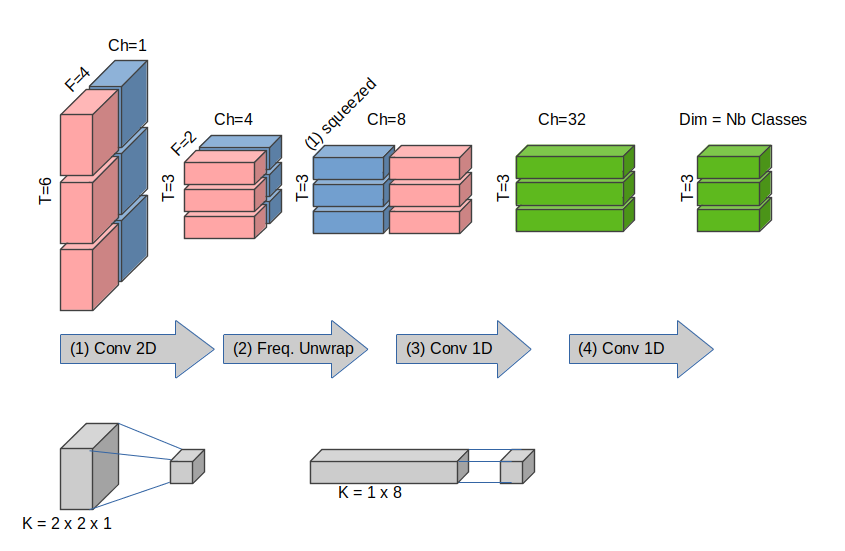}}
		\caption{Differential processing of Time (T) and Frequency (F) by SIMP-FU models illustrated with a minimal example. (1) The input array goes through 2D convolution, T and F are down-sampled by 2x2 while Channel (Ch) is increased from 1 to 4. Convolution regions do not overlap here for clarity. (2) Frequency is unwrapped into the Channel dimension. (3) 1D convolution, Nb Ch arbitrarily set to 32. (4) 1D convolution which maps the 32 channels to output (20 classes). The 1D convolutions use kernel size along time = 1, which is equivalent to fully-connected layers with weights shared across time. Any output unit is connected to all frequencies but only to one of the 3 input time-blocks.}
		\label{fig:explainfu}
	\end{figure}

\begin{figure}[H]
		\fbox{\includegraphics[width=1.0\textwidth]{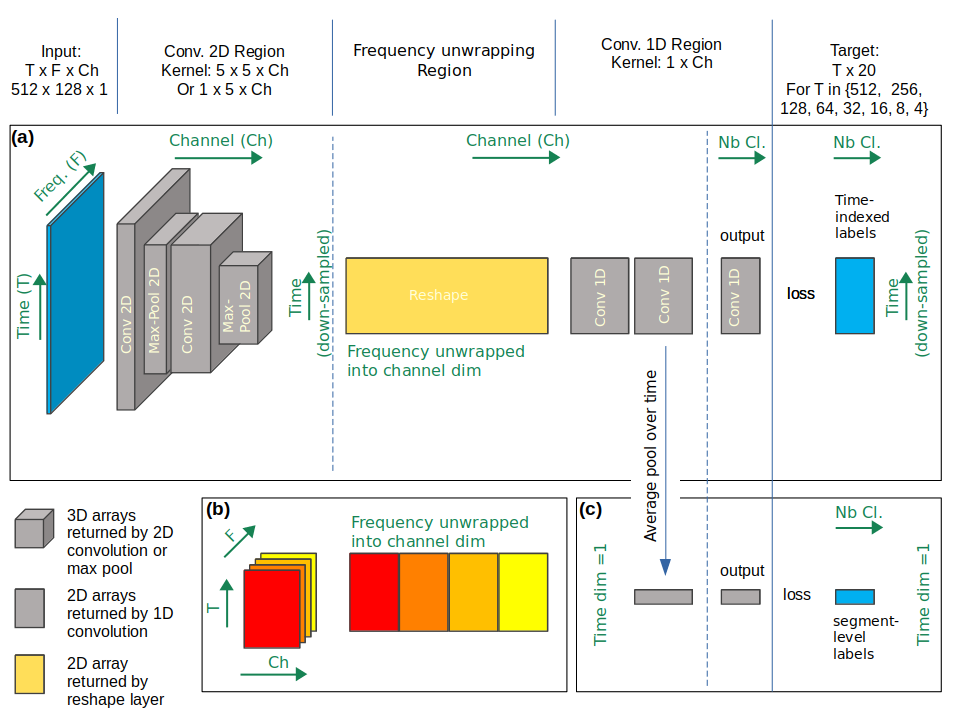}}
		\caption{Overview of the SIMP-FU architectures. Only two 2D convolutional blocks shown here, but between 0 and 7 were used in the actual models. (a) Full schematics of a SIMP-FU CNN with time-indexed labels. (b) Illustration of frequency unwrapping. (c) Alternative output for segment-level labels. Grey blocks represent the output arrays of the layers}
		\label{fig:cnnshema}
	\end{figure}

	We defined a family of CNNs which take spectrograms as input (time x frequency) and return time-indexed, multi-label outputs (time x class). In this study, all models had a fixed input size (512 x 128) while the output had variable time resolutions ranging from 1 (full pooling) to 512 time bins (no pooling). The number of classes was always 20. Any output unit is connected to all 128 frequency bins of the input spectrogram but only to a local region of its time bins. This region is referred to as the MRF hereafter. This was achieved through a Frequency Unwrapping (FU) layer which is explained below and illustrated with a minimal example in Figure \ref{fig:explainfu}. A full description of SIMP-FU models is provided in Figure \ref{fig:cnnshema}. The MRF over time of an output unit was inversely related to the output time resolution (See Table 4).
	
	These architectures allowed us to manipulate the MRF over time. A specific architecture is called a ‘model’ hereafter. Detailed descriptions of a few models are given in Tables 2 and 3 and a summary of all models in Table 4. All CNNs used batch normalization in the convolutional layers and ReLU as transfer functions, sigmoid transfer was used only in the output layer. To avoid an explosion in number of weights, linear increment in channel depth was used in most experiments (A,B,C,D). All models have 3 main regions (see Figure \ref{fig:cnnshema}) which are described hereafter.
	
	\textbf{Conv2D Region} This region was inspired by the VGG architecture \cite{ simonyan2015deep}. It is basically a sequence of convolutional blocks with receptive field 5x5 (3x3 and 3x3) followed by 2x2 max-pooling. The number of convolutional blocks varied from 0 to 7 (Tables 2 and 3). In Figure \ref{fig:cnnshema} only two blocks are depicted. 
	
	\textbf{Frequency Unwrapping Region} The output of the last max-pooling layer is reshaped from 3D to 2D by unwrapping the frequency dimensions into the channel dimension, while the time dimension is conserved (Figure \ref{fig:explainfu} and  \ref{fig:cnnshema} Box b). This has an important impact on the CNN's overall connectivity: Any output unit is connected to all frequency bins of the input spectrogram but only to a local subset of its time bins.
	
	\textbf{Conv1D Region} The output of the Frequency Unwrapping Region was fed into two consecutive 1D convolution layers to allow further nonlinear combination of the channel dimension. These two layers were fixed to 256 channels in all experiments. The convolutional kernel size was fixed to 1 to avoid additional increment of the MRF in time which by design is meant to be controlled only in the Conv2D Region. Finally, the output layer is a 1D convolution (also kernel size = 1) with height equal to the number of time bins and depth equal to number of classes (Figure \ref{fig:cnnshema}, right). Sigmoid transfer was used in the output layer to implement multi-label classification (multiple classes can be present simultaneously). At training time, binary cross-entropy loss was used.
	
	\subsection{Experiments with SIMP-FU CNNs}

	In 5 series of experiments (A to E) we assessed several SIMP-FU architectures. Each series had 8 models covering several values of MRF. See Table 4 for details.
	{\bf (A)} In a first series (A00-A07) the objective was to modify the MRF in time while keeping the total network depth constant. This was achieved by setting kernel and max-pooling in some layers to 1 along the time dimensions as shown in Table 2. At the one end, model A00 used 7 standard convolutional blocks (5x5) resulting in a large MRF of 636 time bins (12.4 sec). MRF can be larger than the input array because padding was used in the convolutional layers. At the other end, Model A07 used only 1x5 convolutional blocks which do not act on the time dimension and the output has 512 time bins (same as the input) and an MRF size of 1 bin (0.02 second).  
	{\bf (B)} In a second series (B00-B07) we used same MRF values as in the first series while making the models shallower. The 1x5 convolutional blocks were removed (Table 3). The B models have a lower level of nonlinear combinations along the frequency dimension in the Conv2D region. {\bf (C)} In a third series (C00-C07) we assessed the usage of simpler segment-level labels because this would reduce the annotation workload. The models from B were modified by including average pooling over time just before the output (Figure \ref{fig:cnnshema}, Box C). In this form they return exactly one value per class and per segment. In a way, the task of identifying the relevant time region is handed over from the human labeler to the training procedure.
	
	The models from A,B, and C have linear layer depth increment (32, 64, 96, 128, 160, 192, 224).
	{\bf (D)} In a fourth series (D00-D07), models were as in B but with depth multiplied by two (64, 128, 192, 256, 320, 384, 448). 
	{\bf (E)} In the fifth series (E00-E07), models were as in B but with exponential increment of depth (32, 64, 128, 256, 512, 1024, 2048). A few models from D and E are equal to models from B and they are marked with an asterisk in Table 4. 

	\subsection{Experiments with image CNNs}
	
	Two of the smaller CNNs developed for image classification were adapted for our setting (Mobilenet V2 \cite{howard2017mobilenets,sandler2018mobilenetv2}, Densenet 121 \cite{huang2017densely}). These architectures have a pooling factor of 32 at the output of the convolutional blocks. The input 512 x 128 is mapped to an output of 16 x 4. These models natively include global average pooling over the time and frequency dimensions. A first baseline experiment aimed at modifying the models minimally (DEN00, MOB00). The last layer (softmax) was replaced by a sigmoid layer to implement multi-label classification as in the SIMP-FU models.  In a second experiment (DEN01, MOB01), the output of their convolutional region was piped into the Frequency Unwrapping and Conv1D blocks of our models (i.e. Conv2D region in Figure \ref{fig:cnnshema} was replaced by the Conv2D region of the image CNNs). Thereby the 4 frequency bins were mapped to separate channels in the Conv1D part while the time resolution at the output was 16. These two models are most comparable to SIMP-FU models of type '-02' (Table 4) which also have an output time resolution of 16.   
	
	\subsection{Training procedure }
	
	The training data was up-sampled on a per class basis to achieve a higher balance across the 20 classes (Table 1). Training-time data augmentation of the spectrograms was applied with the following random distortions: (1) limited range frequency shift, (2) time-shift, (3) mixing of any two randomly selected segments from the training set, including labels. Thereby xeno-canto sounds were often mixed with Mamirauá background noise. One epoch was defined as a training cycle over the up-sampled datasets (14670 segments). SIMP-FU models were trained for 11 epochs, and training was replicated 4 times with different random initialization of the weights and random distortion from data augmentation. The Adam optimizer \cite{kingma2017adam} was used with initial learning rate = 0.001 and learning rate decay = 0.001. Binary cross-entropy was used as loss function. A batch size of 32 was used. The same procedure was used for models DEN and MOB but with a batch size of 16 and 18 epochs. When adequate, the binary labels were down-sampled via block-wise max pooling to match the output’s time resolution. CNNs were implemented with TensorFlow 2.7.0 and trained on an GPU (NVIDIA GeForce RTX 2060).
	
	\subsection{Metric of classification performance}
	
	Models returned predictions as a matrix with N = 20 classes and time resolution between 1 and 512 time bins (Figure \ref{fig:cnnshema}). Test performance was computed at the level of segments. Predictions were summarized as the mean over time for each class to obtain 20 values in [0,1]. For each of the 20 classes, the Average Precision (AP) and Area Under the ROC Curve (AUC) were obtained with the 5635 test segments. As overall performance metrics, the unweighted mean over the per-class metrics (macro mean) was reported for both AP and AUC. Per-class metrics were also reported separately. In our test set, the abundance was very variable depending on the class (Table 1). AUC is insensitive to differing abundance of the classes while AP is sensitive in a way that frequent sound tend to get higher values. AP values are therefore more difficult to compare across classes. As a reference, a classifier that predicts at random gets AUC=~0.5 and AP equals the relative proportion of calls, a hypothetical perfect classifier gets AUC=1.0 and AP=1.0. These two metrics were chosen because they quantify the ability to separate the classes independently of a particular threshold. Methods to choose a threshold were not addressed because it was not the primary focus of this study.
	
	\subsection{Assessment of inference speed}
	
	In a deployment scenario, data will typically be processed in segments of fixed length. A high level of time coverage can be achieved with consecutive segments but this is also costly in terms of computation. This more costly scenario was assessed here. For real-time processing, inference must be faster than the time elapsed between two segments. Here, faster than 10 seconds. For GPU-powered batch processing of large collections of recorded data, inference should be orders of magnitude faster than total data duration. Inference speed was reported as the processing factor (the segment length divided by its inference time). This metric informs on inference time relative to a segment's duration; larger value means faster inference. On compact low-cost devices (Raspberry Pi), segments were processed one at a time with a single thread to mimic their sequential acquisition. On GPU-powered machines, inference was performed by batches of 32 segments.   
	
	\section{Results}
	
	\subsection{Impact of receptive field}
	
\begin{figure}[H]
		\fbox{\includegraphics[width=1.0\textwidth]{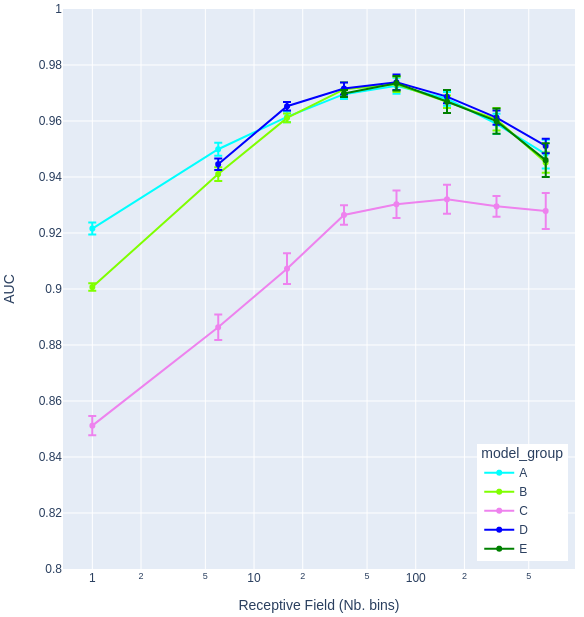} }
		\caption{Macro AUC over all 20 classes for all SIMP-FU models plotted against MRF. Error bars are standard deviation over several training runs. Horizontal axis is logarithmic an goes from 1 to 636 time bins. See Table 4 to convert bins to seconds}%
		\label{fig:curvesauc}
	\end{figure}

\begin{figure}[H]
		\fbox{\includegraphics[width=1.0\textwidth]{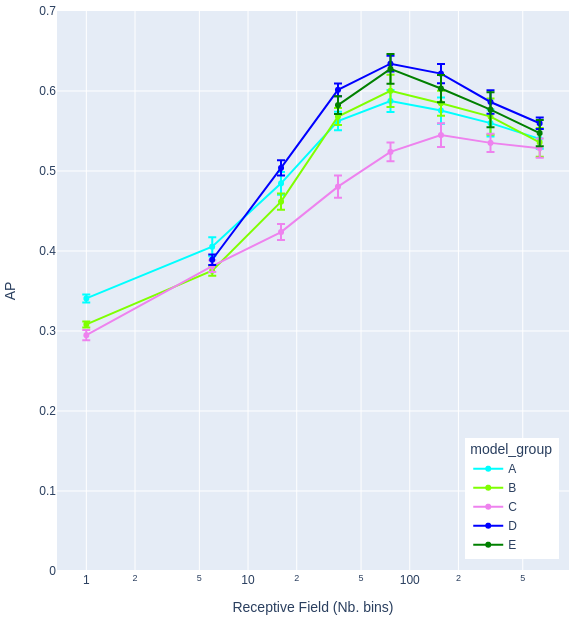} }
		\caption{Macro AP over all 20 classes for all SIMP-FU models plotted against MRF. Error bars are standard deviation over several training runs. Horizontal axis is logarithmic an goes from 1 to 636 time bins. See Table 4 to convert bins to seconds}
		\label{fig:curvesavep}
	\end{figure}

\begin{figure}[H]
		\fbox{\includegraphics[width=1.1\textwidth]{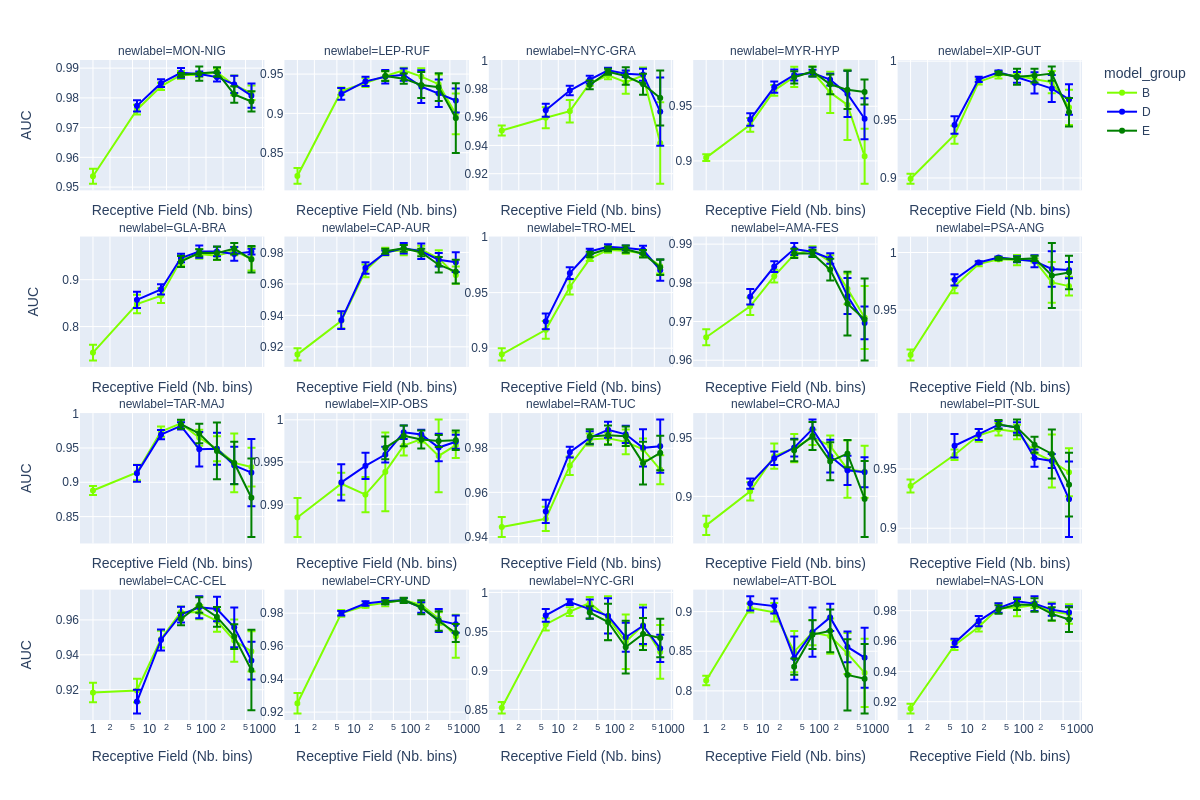} }
		\caption{AUC plotted against MRF separately for each class. Vertical scale independent in each sub-plot to highlight the pattern within each class. For comparison of absolute performance see Figure 9}%
		\label{fig:detailauc}
	\end{figure}

\begin{figure}[H]
		\fbox{\includegraphics[width=1.1\textwidth]{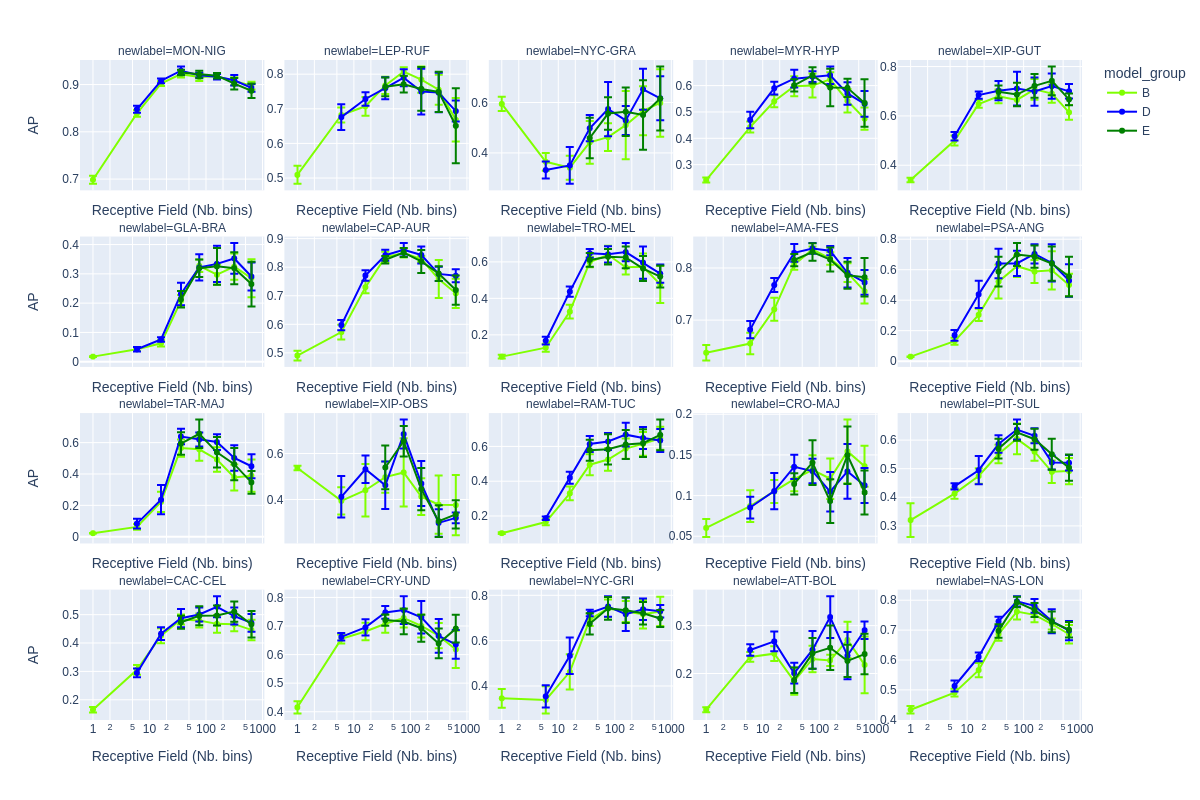} }
		\caption{AP plotted against MRF separately for each class. Vertical scale independent in each sub-plot to highlight the pattern within each class. For comparison of absolute performance see Figure 9}%
		\label{fig:detailavep}
	\end{figure}

\begin{figure}
	\centering
		\begin{subfigure}[b]{1.0\textwidth}
			\centering
			\fbox{\includegraphics[width=1.0\textwidth]{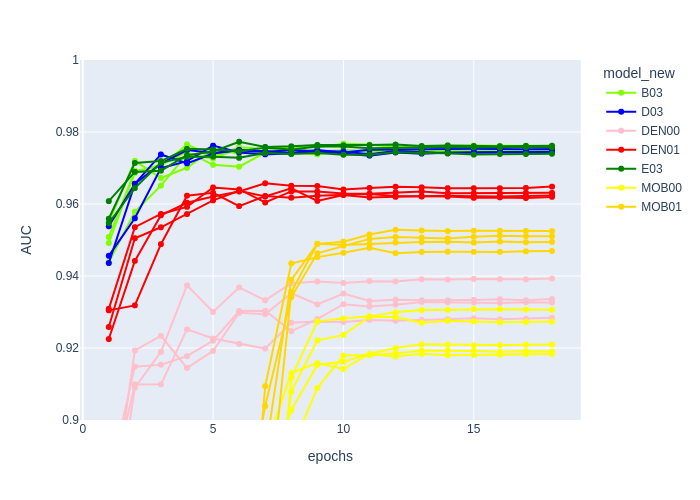}}
		\end{subfigure}
		\begin{subfigure}[b]{1.0\textwidth}
			\centering
			\fbox{\includegraphics[width=1.0\textwidth]{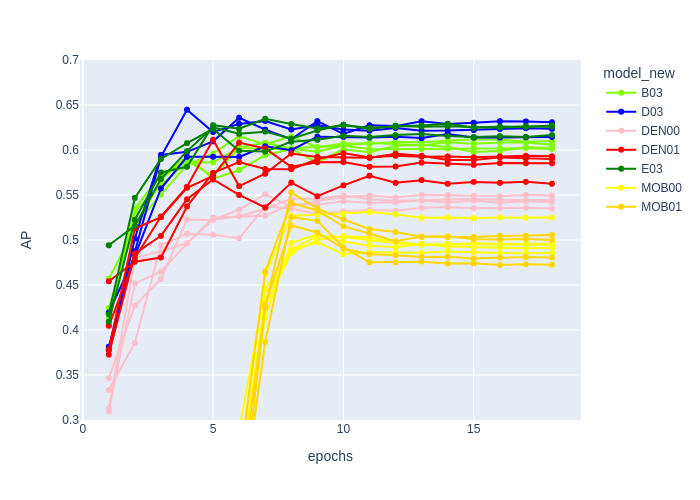}}
		\end{subfigure}
	\caption{Performance vs epoch for the three best performing SIMP-FU models and the image CNN architectures}
	\label{fig:resuauctime}
	\end{figure}

	Models using time-indexed labels (A,B,D,E) performed considerably better than their segment-level counterparts (C). Within the group of time-indexed models, the primary driver of performance was the MRF size. Highest overall performances were achieved at intermediate MRF values of 76 time bins (1.48 sec). There was a drop in performance as MRF became larger or smaller. See Figures \ref{fig:curvesauc} and \ref{fig:curvesavep}. The performance of models in groups A and B was almost equal except for the two smallest MRF values of 1 and 6. Models with deeper convolutional blocks (D,E) performed better than the shallower models from groups B. Noticeably, on a per-class basis, the same association between MRF and classification performance was observed in the majority of classes (Figures \ref{fig:detailauc} and \ref{fig:detailavep}). FU-modified image CNNs (DEN01, MOB01) performed better than their respective unmodified counterparts (DEN00, MOB00). DEN01 performed best among image CNNs but lower than the best SIMP-FU models (Figure \ref{fig:resuauctime}).
	
	\subsection{Absolute classification performance}
	
\begin{figure}
	\centering
		\begin{subfigure}[b]{1.1\textwidth}
			\centering
			\fbox{\includegraphics[width=1.0\textwidth]{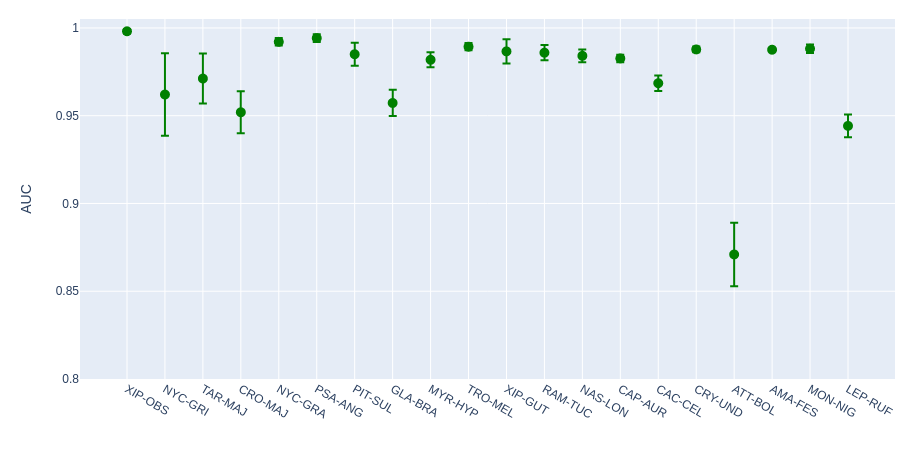}}
		\end{subfigure}
		\begin{subfigure}[b]{1.1\textwidth}
			\centering
			\fbox{\includegraphics[width=1.0\textwidth]{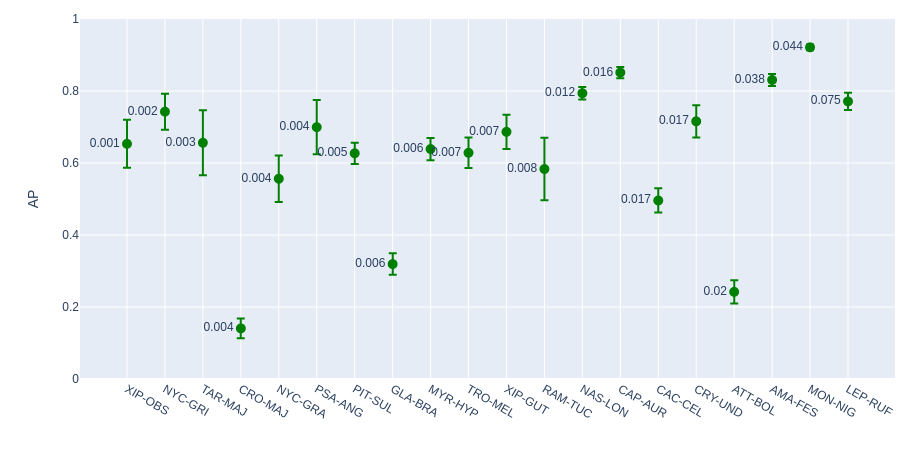}}
		\end{subfigure}
	\caption{Per-class predictive performance of model E03. For AP the proportion of each class in the test set is annotated at each dot. Sound-types ordered left to right from lowest  to highest proportion. Error bars are standard deviation over several training runs.}
	\label{fig:perlabelperf}
	\end{figure}

	In absolute terms, classification performance of some SIMP-FU models was good, e.g. Model E03 with AUC above 0.95 in 18 sound-types and even above 0.98 in 13 sound-types (Figure \ref{fig:perlabelperf} a). Out of 12 sound-types with abundance below 0.01, 10 reached AP above 0.5. Out of 8 other sound-types with  abundance below 0.10, 6 reached AP above 0.5 (Figure \ref{fig:perlabelperf} b). 
	
	\subsection{Inference speed}
	
\begin{figure}[H]
	\subfloat[\centering Compact low-cost device CPU]{
		{\fbox{\includegraphics[width=0.5\textwidth]{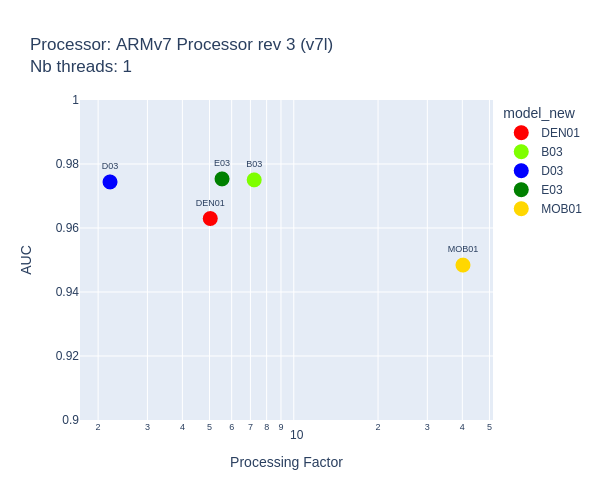} }}}
	\subfloat[\centering High performance GPU processor]{
		{\fbox{\includegraphics[width=0.5\textwidth]{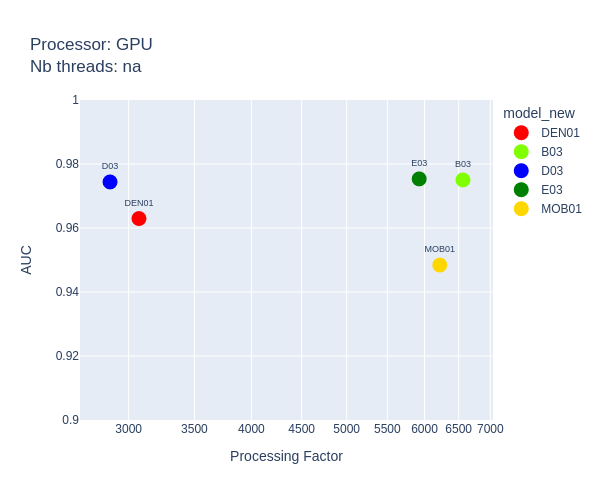} }}}
	\caption{Classification strength (AUC) of subset of models plotted against processing factor from inference on a compact low-cost CPU with a single thread and batch size of 1 (a), and GPU processing with batch size of 32 (b).}
	\label{fig:timingall}
	\end{figure}

	On compact low-cost hardware (Raspberry Pi) using only one thread, D03 evaluated approx. two times faster than data acquisition, E03 five times, and B03 seven times faster (Figure \ref{fig:timingall} a). In terms of speed, MOB01 was the fastest and DEN01 was in the same range as E03. However, the classification performance of MOB01 and DEN01 was lower than the SIMP-FU models. On GPU D03 evaluated approx. 3000 times faster than data acquisition while B03 and E03 evaluated approx. 6000 times faster (Figure \ref{fig:timingall} b). 
	
	\section{Discussion}
	
	\subsection{Segment-level vs time-indexed labels}
	
	In the current setting of moderate number of training examples, time-indexed models were superior to segment-level models (Groups A,B,D,E vs. C in Figures \ref{fig:curvesauc} and \ref{fig:curvesavep}).The better generalization performance of time-indexed models can be explained by at least two mechanisms: (1) Many sound-types are shorter than the segments and time-indexed labeling explicitly marks the surrounding noise as background such that the time-indexed models can learn to discriminate the target sound from its direct surrounding. This reduces the risk of wrongly learning background patterns that co-occurred with target sounds in the training set. (2) In data augmentation, the mixing of 2 segments with time-indexed labels allowed to create diverse situations where two different sound-types were in the same segment but generally only partially overlapped. It thereby increased the diversity of the training set by creating challenging situations that were infrequent in the un-augmented data. Time-indexed annotation with bounding boxes was more work intensive than just tagging the presence of sounds at the level of segments. Our results show that time-indexed annotation was worth the effort.
	 
	\subsection{Receptive field}
	
	The best performing MRF of 1.48 seconds is in the same range as the duration of 8 fixed duration sounds (FD2, FD3) but shorter than 4 sounds (FD1) (Table 1). For all 8 variable duration (VD) sounds an MRF of 1.48 seconds seems sufficient to capture a few basic units of the repeating pattern (see figures in Appendix). In summary, the MRF of 1.48 is adequate to capture the full informative patterns of 16 sounds but it can only capture partial information in 4 sounds of the group FD1. Due to the segmented nature of the data, training examples of longer sounds (FD1) were often truncated and due to overlap with other sounds they were often partially masked. Hence, many training examples of FD1 were de-facto shorter than the full duration of the sounds. This may have prevented the CNNs from learning longer patterns. Working with longer segments, e.g. one minute, should be assessed in the future. A key lesson from our study is that the duration of the receptive field matters. 
	
	Our models taken individually impose a single MRF on all 20 sound-types. In our dataset short sounds or repetitive sounds with a short base pattern were predominant and pointed us to  relatively short value of MRF. The current study design probably prevented us from finding optimal models for the longer sounds. In the future we propose to train separate models for shorter and longer sounds or even dedicated models for each class.
	
	\subsection{Model complexity vs performance}
	
	Within each depth group, the best performing models were D03, E03, and B03 with 2.4, 1.7 and 0.7 millions trainable weights. These three models were smaller and achieved better classification performance on an acoustic classification task compared to the best image CNN, DEN01, with 8.1 M weights. Generally, the patterns found in spectrogram data are considerably simpler than patterns in photograph images. The expressive power of SIMP-FU models seems sufficient to handle the complexity of acoustic patterns from the 20 sound-types considered here. Many animal sounds are expected to fall into this category of complexity. We propose that under the constraint of moderate-sized training sets, simple CNNs such as SIMP-FU models can be used for bioacoustic monitoring. 
	
	\subsection{Inference speed}
	
	While model D03 performed best in terms of predictive performance, it was only two times faster than acquisition rate on low-cost hardware. Considering that other processes need to run on deployed sensors (e.g. spectrogram computation, equalization and normalization), this is probably too slow. Models B03 and E03 were close in predictive performance and evaluated 5 and 7 times faster, which leaves sufficient resources for other processes. Model E03 seems to offer a good balance between speed and classification performance; it has 32 channels in the first layers vs 64 for model D03. This considerable improvement in speed is because early convolutional layers are the most time consuming as they act on the 'full-sized' spectrogram. To further optimize the trade-off between speed and classification performance, future research should also focus on the first layer and on minimizing the size of the spectrogram.
	
	GPUs can be used for batch processing of large data collections. On GPU the faster well performing models (B03 and E03) ran twice as fast as the larger model (D03). This is a considerable difference, especially considering that access to GPU powered machine is limited in institutions with low financial resources. E03 seems to offer an optimal balance between speed and predictive performance also on GPU. 
	
	\subsection{Side note}
	
	A full data processing pipeline for long-term bioacoustic monitoring must include more steps that were not addressed here, such as: (1) smart choice of decision thresholds based on expected call rate, (2) handling of interference sounds such as heavy rain/wind, plane and ship engine, human speech, saturation due to the animal very close to microphone, (3) aggregated decision values at periods longer than 10 seconds, e.g. 5 minutes. 
	
	\subsection{Lookout}
	
	Large scale bioacoustic monitoring will need to include many animal species from diverse groups. However, only moderate or even small datasets can be expected to be available for training in general because labeling is a fundamental bottleneck in ML projects. Our models achieved good predictive performance by learning from a moderate-sized training set with less than 100 examples in most classes. Such small models could be an important contribution to realize large scale bioacoustic monitoring. Complementary to large data-greedy DNNs that handle all classes at once, we envision ensembles of smaller CNNs with well defined properties like the MRF, where each CNN is adapted for a particular groups of sounds. We used a single input frequency range for all classes (100-5000 Hz). In the future it could be advantageous to use dedicated CNNs for several smaller frequency ranges. 
	
	\section{Conclusion}
	
	\begin{enumerate}
		\item 
		The duration of the receptive field was identified as a critical aspect that conditions the generalization performance of CNNs used for bioacoustic classification.

		\item Models that can leverage time-indexed labels during training achieved considerably better generalization performance compared to models that use labels at the level of 10 seconds.     
		
	 	\item For bioacoustic classification, comparison of models covering a wide range of capacities showed that smaller and faster models had better generalization performance. 
		
		\item Our results confirm the feasibility of deploying small CNNs with good generalization performance on a large number of compact low-cost devices.
	\end{enumerate}
	
	\section{Acknowledgements}\
	
	We would like to thank Guilherme Costa Alvarenga, Kotaro Tanaka, Monika Kosecka, Pedro Nassar, Valciney Martins, and Wezddy Del Toro Orozco for their valuable help in defining the sound classes and during the early phase of labelling the sound examples. We are very thankful to the Gordon and Betty Moore Foundation (www.moore.org) who funded Project PROVIDENCE and helped set the basis for the current work. 
	
	\newpage
	
	\section{Tables}\

	\begin{table}[H]
		\scalebox{0.83}{\begin{filecontents*}{tabs/m01_b_descriptives_short.csv}
Species,Short name, Group,Description,Train,Train,Test,Test  
----- Fixed duration ,,,--- Approx duration,N,N ups.,N,Prop
Attila bolivianus ,ATT-BOL,FD1,3+ sec,32,224,115,0.02
Taraba major,TAR-MAJ,FD1,3+ sec,40,200,15,0.003
Nyctibius griseus,NYC-GRI,FD1,3+ sec,39,234,10,0.002
Xiphorhynchus guttatus,XIP-GUT,FD1,3+ sec,41,205,40,0.007
Myrmelastes hyperythrus,MYR-HYP,FD2,1 – 3 sec,40,224,33,0.006
Xiphorhynchus obsoletus,XIP-OBS,FD2,1 – 3 sec,33,231,8,0.001
Crypturellus undulatus,CRY-UND,FD2,1 – 3 sec,116,251,96,0.017
Nasica longirostris,NAS-LON,FD3,0.5 – 1 sec,64,297,65,0.012
Nyctibius grandis,NYC-GRA,FD3,0.5 – 1 sec,60,243,22,0.004
Psarocolius angustifrons,PSA-ANG,FD3,0.5 – 1 sec,38,228,24,0.004
Pitangus sulphuratus,PIT-SUL,FD3,0.5 – 1 sec,73,222,26,0.005
Leptotila rufaxila,LEP-RUF,FD3,0.5 – 1 sec,62,269,420,0.075
----- Variable duration,,,--- Repet. interval,,,,
Ramphastos tucanus,RAM-TUC,VD1,0.5 sec,94,323,44,0.008
Trogon melanurus,TRO-MEL,VD1,0.7 sec,93,331,38,0.007
Capito aurovirens,CAP-AUR,VD1,0.5 sec,313,449,90,0.016
Glaucidium brasilianum,GLA-BRA,VD1,0.3 sec,87,265,33,0.006
Amazona festiva,AMA-FES,VD2,irregular,99,305,213,0.038
Cacicus cela,CAC-CEL,VD2,irregular,111,229,95,0.017
Crotophaga major,CRO-MAJ,VD2,irregular,27,216,22,0.004
Monasa nigrifrons,MON-NIG,VD2,irregular,135,300,250,0.044
Total nb of segments,,,,11341,14670,5635,
\end{filecontents*}
\csvautotabular	{tabs/m01_b_descriptives_short.csv}}
		\caption{Grouping of sound-types according to their typical duration and summary of training and test data. FD: Fixed Duration, VD: Variable duration. N: count of segments containing a particular sound-type. N ups: count after re-balancing classes of training set. The total nb of segments was larger than the sum of counts because many segments did not contain any of the 20 sound-types. Proportion of each sound-type in test set is given in last column.}
	\end{table}
	
	\begin{table}[H]
		\scalebox{0.79}{
		\begin{filecontents*}{tabs/m03E1xxarchitectures.csv}
,A00,(Conv)  (mp),A03,(Conv)  (mp),A07,(Conv)  (mp)
input ,512 x 128 x 1,(5x5) (2x2),512 x 128 x 1,(5x5) (2x2),512 x 128 x 1,(1x5) (1x2)
output block 1,256 x 64 x 32,(5x5) (2x2),256 x 64 x 32,(5x5) (2x2),512 x 64 x 32,(1x5) (1x2)
output block 2,128 x 32 x 64,(5x5) (2x2),128 x 32 x 64,(5x5) (2x2),512 x 32 x 64,(1x5) (1x2)
output block 3,64 x 16 x 96,(5x5) (2x2),64 x 16 x 96,(5x5) (2x2),512 x 16 x 96,(1x5) (1x2)
output block 4,32 x 8 x 128,(5x5) (2x2),32 x 8 x 128,(1x5) (1x2),512 x 8 x 128,(1x5) (1x2)
output block 5,16 x 4 x 160,(5x5) (2x2),32 x 4 x 160,(1x5) (1x2),512 x 4 x 160,(1x5) (1x2)
output block 6,8 x 2 x 192,(5x5) (2x2),32 x 2 x 192,(1x5) (1x2),512 x 2 x 192,(1x5) (1x2)
output block 7,4 x 1 x 224,,32 x 1 x 224,,512 x 1 x 224,
freq. unwr.,4 x 224,(1) (na),32 x 224,(1) (na),512 x 224,(1) (na)
output conv1D,4 x 256,(1) (na),32 x 256,(1) (na),512 x 256,(1) (na)
output conv1D,4 x 256,(1) (na),32 x 256,(1) (na),512 x 256,(1) (na)
output T x Cl,4 x 20,,32 x 20,,512 x 20,
---,---,---,---,---,---,---
MRF (time bins) ,636,,76,,1,
MRF (seconds) ,12.42,,1.48,,0.02,
\end{filecontents*}
\csvautotabular{tabs/m03E1xxarchitectures.csv}
		}
		\caption{Description of array and convolutional kernel sizes in 3 models from the A group. mp: max pooling, T: Time, F: Frequency, Ch: Channel, Cl: Classes. Before frequency unwrapping array size given as T x F x Ch and after as T x Ch}
	\end{table}
	
	\begin{table}[H]
		\scalebox{0.78}{
		\begin{filecontents*}{tabs/m04E3xxarchitectures.csv}
,B00,(Conv)  (mp),B03,(Conv)  (mp),B07,(Conv)  (mp)
input ,512 x 128 x 1,(5x5) (2x2),512 x 128 x 1,(5x5) (2x2),512 x 128 x 1,
output block 1,256 x 64 x 32,(5x5) (2x2),256 x 64 x 32,(5x5) (2x2),,
output block 2,128 x 32 x 64,(5x5) (2x2),128 x 32 x 64,(5x5) (2x2),,
output block 3,64 x 16 x 96,(5x5) (2x2),64 x 16 x 96,(5x5) (2x2),,
output block 4,32 x 8 x 128,(5x5) (2x2),32 x 8 x 128,,,
output block 5,16 x 4 x 160,(5x5) (2x2),,,,
output block 6,8 x 2 x 192,(5x5) (2x2),,,,
output block 7,4 x 1 x 224,,,,,
freq. unwr.,4 x 224,(1) (na),32 x 1024,(1) (na),512 x 128,(1) (na)
output conv1D,4 x 256,(1) (na),32 x 256,(1) (na),512 x 256,(1) (na)
output conv1D,4 x 256,(1) (na),32 x 256,(1) (na),512 x 256,(1) (na)
output T x Cl,4 x 20,,32 x 20,,512 x 20,
---,---,---,---,---,---,---
MRF (time bins) ,636,,76,,1,
MRF (seconds) ,12.42,,1.48,,0.02,

\end{filecontents*}
\csvautotabular{tabs/m04E3xxarchitectures.csv}
		}
		\caption{Description of array and convolutional kernel sizes in 3 models from the B group. Letter-Codes: see Table 2}
	\end{table}
		
	\begin{table}[H]
		\scalebox{0.62}{	
		\begin{filecontents*}{tabs/m05_final_archs_wide.csv}
Model group,Model Number,--00,--01,--02,--03,--04,--05,--06,--07
,--- Nb conv. Blocks,7,7,7,7,7,7,7,7
A,32-lin,2457206,1898102,1492598,1216118,1044086,951926,915062,908726
,--- Nb conv. Blocks,7,6,5,4,3,2,1,0
B,32-lin,2457206,1658230,1114422,797110,669430,661750 (*2),606134 (*3),104980 (*4)
C (*1),32-lin,2457206,1658230,1114422,797110,669430,661750,606134,104980
D,64-lin,9487062,6211798,3907670,2443606,1672150,1380310,1158486,104980 (*4)
E,32-exp,76105334,19471990,5311094,1769590,883574,661750 (*2),606134 (*3),104980 (*4)
---,---,---,---,---,---,---,---,---,---
,Time resol. at output (*1),4,8,16,32,64,128,256,512
,MRF (time bins),636,316,156,76,36,16,6,1
,MRF (seconds),12.42,6.17,3.05,1.48,0.7,0.31,0.12,0.02

\end{filecontents*}
\csvautotabular{tabs/m05_final_archs_wide.csv}	
		}
		\caption{Summary of all SIMP-FU models with number of trainable weights (upper part of table). Within each model group there were 8 models (00 to 07) with different output shape and MRF (lower part of table). 
			*1 Model C as B but with average pooling over time in the layer preceding output. In C, time resolution is value before average pooling. 
			*2 Model B05 = E05; *3 B06 = E06; *4 B07=E07=D07; only B05, B06, and B07 reported in figures.
			Increment in channel depth:
			32-lin: 32, 64, 96, 128, 160, 192, 224.
			64-lin: 64, 128, 192, 256, 320, 384, 448.
			32-exp: 32, 64, 128, 256, 512, 1024, 2048.	
			}
	\end{table}		

	\begin{table}[H]
	\scalebox{0.9}{	
		\begin{filecontents*}{tabs/m06imagearchitectures.csv}
model,based on ,Modified,trainable weights
DEN00,Densenet 121,Sigmoid output,6968084
DEN01,Densenet 121,GA repl. by FU,8068372
MOB00,Mobilenet V2,Sigmoid output,2248916
MOB01,Mobilenet V2,GA repl. by FU,3606228
\end{filecontents*}
\csvautotabular{tabs/m06imagearchitectures.csv}
		}
	\caption{Overview of modified image CNNs used in this study.
	GAP: Global Average Pooling.
	FU: Frequency Unwrapping
	}
	\end{table}		

	\newpage
	
	\section{Figures}\
	
	\newpage
	
	\section{References}\
	
	\printbibliography

@article{geirhos2020shortcut,
	title={Shortcut learning in deep neural networks},
	author={Geirhos, Robert and Jacobsen, J{\"o}rn-Henrik and Michaelis,
	Claudio and Zemel, Richard and Brendel, Wieland and Bethge, Matthias and
	Wichmann, Felix A},
	journal={Nature Machine Intelligence},
	volume={2},
	number={11},
	pages={665--673},
	year={2020},
	publisher={Nature Publishing Group}
}

@article{araujo2019computing,
	author = {Araujo, André and Norris, Wade and Sim, Jack},
	title = {Computing Receptive Fields of Convolutional Neural Networks},
	journal = {Distill},
	year = {2019},
	doi = {10.23915/distill.00021}
}

@misc{stowell2021computational,
	title={Computational bioacoustics with deep learning: a review and roadmap}, 
	author={Dan Stowell},
	year={2021},
	eprint={2112.06725},
	archivePrefix={arXiv},
	primaryClass={cs.SD}
}

@article{lebien2020pipeline,
	title={A pipeline for identification of bird and frog species in tropical soundscape recordings using a convolutional neural network},
	author={LeBien, Jack and Zhong, Ming and Campos-Cerqueira, Marconi and Velev, Julian P and Dodhia, Rahul and Ferres, Juan Lavista and Aide, T Mitchell},
	journal={Ecological Informatics},
	volume={59},
	pages={101113},
	year={2020},
	publisher={Elsevier}
}

@misc{damour2020underspecification,
	title={Underspecification Presents Challenges for Credibility in Modern Machine Learning}, 
	author={Alexander D'Amour and Katherine Heller and Dan Moldovan and Ben Adlam and Babak Alipanahi and Alex Beutel and Christina Chen and Jonathan Deaton and Jacob Eisenstein and Matthew D. Hoffman and Farhad Hormozdiari and Neil Houlsby and Shaobo Hou and Ghassen Jerfel and Alan Karthikesalingam and Mario Lucic and Yian Ma and Cory McLean and Diana Mincu and Akinori Mitani and Andrea Montanari and Zachary Nado and Vivek Natarajan and Christopher Nielson and Thomas F. Osborne and Rajiv Raman and Kim Ramasamy and Rory Sayres and Jessica Schrouff and Martin Seneviratne and Shannon Sequeira and Harini Suresh and Victor Veitch and Max Vladymyrov and Xuezhi Wang and Kellie Webster and Steve Yadlowsky and Taedong Yun and Xiaohua Zhai and D. Sculley},
	year={2020},
	eprint={2011.03395},
	archivePrefix={arXiv},
	primaryClass={cs.LG}
}

@article{simonyan2015deep,
	title={Very Deep Convolutional Networks for Large-Scale Image Recognition}, 
	author={Karen Simonyan and Andrew Zisserman},
	year={2015},
	eprint={1409.1556},
	archivePrefix={arXiv},
	primaryClass={cs.CV}
}

@inproceedings{bittle2013review,
	title={A review of current marine mammal detection and classification algorithms for use in automated passive acoustic monitoring},
	author={Bittle, Michael and Duncan, Alec},
	booktitle={Proceedings of Acoustics},
	volume={2013},
	year={2013},
	organization={Citeseer}
}

@article{malfante2018automatic,
	title={Automatic fish sounds classification},
	author={Malfante, Marielle and Mars, J{\'e}r{\^o}me I and Dalla Mura, Mauro and Gervaise, C{\'e}dric},
	journal={The Journal of the Acoustical Society of America},
	volume={143},
	number={5},
	pages={2834--2846},
	year={2018},
	publisher={Acoustical Society of America}
}

@article{priyadarshani2018automated,
	title={Automated birdsong recognition in complex acoustic environments: a review},
	author={Priyadarshani, Nirosha and Marsland, Stephen and Castro, Isabel},
	journal={Journal of Avian Biology},
	volume={49},
	number={5},
	pages={jav--01447},
	year={2018},
	publisher={Wiley Online Library}
}

@article{xie2018frog,
	title={Frog call classification: a survey},
	author={Xie, Jie and Towsey, Michael and Zhang, Jinglan and Roe, Paul},
	journal={Artificial Intelligence Review},
	volume={49},
	number={3},
	pages={375--391},
	year={2018},
	publisher={Springer}
}

@article{gibb2019emerging,
	title={Emerging opportunities and challenges for passive acoustics in ecological assessment and monitoring},
	author={Gibb, Rory and Browning, Ella and Glover-Kapfer, Paul and Jones, Kate E},
	journal={Methods in Ecology and Evolution},
	volume={10},
	number={2},
	pages={169--185},
	year={2019},
	publisher={Wiley Online Library}
}

@article{shiu2020deep,
	title={Deep neural networks for automated detection of marine mammal species},
	author={Shiu, Yu and Palmer, KJ and Roch, Marie A and Fleishman, Erica and Liu, Xiaobai and Nosal, Eva-Marie and Helble, Tyler and Cholewiak, Danielle and Gillespie, Douglas and Klinck, Holger},
	journal={Scientific reports},
	volume={10},
	number={1},
	pages={1--12},
	year={2020},
	publisher={Nature Publishing Group}
}

@inproceedings{grill2017two,
	title={Two convolutional neural networks for bird detection in audio signals},
	author={Grill, Thomas and Schl{\"u}ter, Jan},
	booktitle={2017 25th European Signal Processing Conference (EUSIPCO)},
	pages={1764--1768},
	year={2017},
	organization={IEEE}
}

@incollection{kahl2017large,
	title={Large-Scale Bird Sound Classification using Convolutional Neural Networks.},
	author={Kahl, Stefan and Wilhelm-Stein, Thomas and Hussein, Hussein and Klinck, Holger and Kowerko, Danny and Ritter, Marc and Eibl, Maximilian},
	booktitle={CLEF (working notes)},
	volume={1866},
	year={2017}
}

@inproceedings{salamon2017fusing,
	title={Fusing shallow and deep learning for bioacoustic bird species classification},
	author={Salamon, Justin and Bello, Juan Pablo and Farnsworth, Andrew and Kelling, Steve},
	booktitle={2017 IEEE international conference on acoustics, speech and signal processing (ICASSP)},
	pages={141--145},
	year={2017},
	organization={IEEE}
}

@incollection{sevilla2017audio,
	title={Audio Bird Classification with Inception-v4 extended with Time and Time-Frequency Attention Mechanisms.},
	author={Sevilla, Antoine and Glotin, Herv{\'e}},
	booktitle={CLEF (Working Notes)},
	volume={1866},
	year={2017}
}

@inproceedings{incze2018bird,
	title={Bird sound recognition using a convolutional neural network},
	author={Incze, Agnes and Jancs{\'o}, Henrietta-Bernadett and Szil{\'a}gyi, Zolt{\'a}n and Farkas, Attila and Sulyok, Csaba},
	booktitle={2018 IEEE 16th International Symposium on Intelligent Systems and Informatics (SISY)},
	pages={000295--000300},
	year={2018},
	organization={IEEE}
}

@article{lasseck2018audio,
	title={Audio-based Bird Species Identification with Deep Convolutional Neural Networks.},
	author={Lasseck, Mario},
	journal={CLEF (Working Notes)},
	volume={2125},
	year={2018}
}

@inproceedings{liu2018classification,
	title={Classification of cetacean whistles based on convolutional neural network},
	author={Liu, Songzuo and Liu, Meng and Wang, Mengjia and Ma, Tianlong and Qing, Xin},
	booktitle={2018 10th International Conference on Wireless Communications and Signal Processing (WCSP)},
	pages={1--5},
	year={2018},
	organization={IEEE}
}

@article{mac2018bat,
	title={Bat detective—Deep learning tools for bat acoustic signal detection},
	author={Mac Aodha, Oisin and Gibb, Rory and Barlow, Kate E and Browning, Ella and Firman, Michael and Freeman, Robin and Harder, Briana and Kinsey, Libby and Mead, Gary R and Newson, Stuart E and others},
	journal={PLoS computational biology},
	volume={14},
	number={3},
	pages={e1005995},
	year={2018},
	publisher={Public Library of Science San Francisco, CA USA}
}

@inproceedings{himawan2018deep,
	title={Deep Learning Techniques for Koala Activity Detection.},
	author={Himawan, Ivan and Towsey, Michael and Law, Bradley and Roe, Paul},
	booktitle={INTERSPEECH},
	pages={2107--2111},
	year={2018}
}

@article{ibrahim2018automatic,
	title={Automatic classification of grouper species by their sounds using deep neural networks},
	author={Ibrahim, Ali K and Zhuang, Hanqi and Ch{\'e}rubin, Laurent M and Sch{\"a}rer-Umpierre, Michelle T and Erdol, Nurgun},
	journal={The Journal of the Acoustical Society of America},
	volume={144},
	number={3},
	pages={EL196--EL202},
	year={2018},
	publisher={Acoustical Society of America}
}

@misc{howard2017mobilenets,
	title={MobileNets: Efficient Convolutional Neural Networks for Mobile Vision Applications}, 
	author={Andrew G. Howard and Menglong Zhu and Bo Chen and Dmitry Kalenichenko and Weijun Wang and Tobias Weyand and Marco Andreetto and Hartwig Adam},
	year={2017},
	eprint={1704.04861},
	archivePrefix={arXiv},
	primaryClass={cs.CV}
}

@inproceedings{sandler2018mobilenetv2,
	title={Mobilenetv2: Inverted residuals and linear bottlenecks},
	author={Sandler, Mark and Howard, Andrew and Zhu, Menglong and Zhmoginov, Andrey and Chen, Liang-Chieh},
	booktitle={Proceedings of the IEEE conference on computer vision and pattern recognition},
	pages={4510--4520},
	year={2018}
}

@inproceedings{huang2017densely,
	title={Densely connected convolutional networks},
	author={Huang, Gao and Liu, Zhuang and Van Der Maaten, Laurens and Weinberger, Kilian Q},
	booktitle={Proceedings of the IEEE conference on computer vision and pattern recognition},
	pages={4700--4708},
	year={2017}
}

@misc{kingma2017adam,
	title={Adam: A Method for Stochastic Optimization}, 
	author={Diederik P. Kingma and Jimmy Ba},
	year={2017},
	eprint={1412.6980},
	archivePrefix={arXiv},
	primaryClass={cs.LG}
}

@inproceedings{luo2016understanding,
	title={Understanding the effective receptive field in deep convolutional neural networks},
	author={Luo, Wenjie and Li, Yujia and Urtasun, Raquel and Zemel, Richard},
	booktitle={Proceedings of the 30th International Conference on Neural Information Processing Systems},
	pages={4905--4913},
	year={2016}
}

	\newpage
	\section{Appendix}\
	
		\textbf{External training set - detailed description} This data was downloaded from the xeno-canto repository (www.xeno-canto.org). Geographic range was limited to Bolivia, Brazil, Ecuador, and Peru. The objective was to provide clear and unambiguous examples of the sound-types. We attempted to download at least 10 MP3 files of at least 600 seconds and took only files tagged with quality A or B whenever possible. Preference was given to shorter files but a file duration of at least 10 s was required. Files with a Non-Derivative (ND) license were excluded. All MP3 files were converted to WAV format and up-sampled to 48000 samples per second. After careful scrutiny of the metadata, we found that this data included 5 out of 547 files (212 out of 13302 seconds) that were recorded in the MSDR independently of the test set and did not overlap in time with the test set. These are highly selected recordings performed or selected by humans, sometimes with directional microphones. The sounds typically exhibit a high SNR. They are not representative of continuous recordings from the wild. While this data contains examples of target sounds, the background noise diversity is probably underrepresented.

	\begin{table}[H]
		\scalebox{0.70}{\begin{filecontents*}{tabs/a01_appendix_xc_download.csv}
sound type,Bolivia,Brazil,Ecuador,Peru,Qual. A,Qual. B,Qual. C, Tot nb files ,Tot length (sec)
Amazona festiva,0,10,5,0,6,7,2,15,537
Attila bolivianus,2,17,0,5,12,12,0,24,641
Cacicus cela,1,12,18,4,24,11,0,35,639
Capito aurovirens,0,4,6,1,1,7,3,11,470
Crotophaga major,2,18,6,1,17,10,0,27,725
Crypturellus undulatus,0,19,4,3,19,7,0,26,666
Glaucidium brasilianum,2,23,4,1,20,10,0,30,663
Leptotila rufaxilla,2,14,6,4,16,10,0,26,682
Monasa nigrifrons,2,19,8,5,19,15,0,34,761
Myrmelastes hyperythrus,3,5,8,10,15,11,0,26,645
Nasica longirostris,1,11,10,2,14,10,0,24,606
Nyctibius grandis,2,9,2,3,3,13,0,16,713
Nyctibius griseus,1,15,6,0,13,9,0,22,643
Pitangus sulphuratus,3,22,11,2,20,18,0,38,663
Psarocolius angustifrons,4,7,26,4,28,13,0,41,731
Ramphastos tucanus,0,17,9,5,23,0,8,31,847
Taraba major,5,21,14,5,26,19,0,45,847
Trogon melanurus,3,8,7,0,15,3,0,18,627
Xiphorhynchus guttatus,4,24,7,5,22,18,0,40,722
Xiphorhynchus obsoletus,4,7,6,1,10,7,1,18,474
All,41,282,163,61,323,210,14,547,13302
\end{filecontents*}
\csvautotabular {tabs/a01_appendix_xc_download.csv}}
		\caption{Details on files downloaded from xeno-canto that were included into the training set.}
	\end{table}

	\begin{table}[H]
		\scalebox{0.62}{\begin{filecontents*}{tabs/m01_b_descriptives_full.csv}
Species,Description ,Duration Group,Additional Metric,Train,Train,Train,Train,Test,Test  
,,,,Period 2,XC,All,Up-sampled,Period 1,Period 1
,,,,N,N,N,N,N,Prop
----- Fixed duration ,,,Approx duration,10242,1099,11341,14670,5635, 
Attila bolivianus ,Song,FD1,3+ sec,1,31,32,224,115,0.02
Taraba major,Song,FD1,3+ sec,0,40,40,200,15,0.003
Nyctibius griseus,Song,FD1,3+ sec,1,38,39,234,10,0.002
Xiphorhynchus guttatus,Song,FD1,3+ sec,9,32,41,205,40,0.007
Myrmelastes hyperythrus,Song,FD2,1 – 3 sec,3,37,40,224,33,0.006
Xiphorhynchus obsoletus,Song,FD2,1 – 3 sec,0,33,33,231,8,0.001
Crypturellus undulatus,Song,FD2,1 – 3 sec,73,43,116,251,96,0.017
Nasica longirostris,Song,FD3,0.5 – 1 sec,36,28,64,297,65,0.012
Nyctibius grandis,Song,FD3,0.5 – 1 sec,8,52,60,243,22,0.004
Psarocolius angustifrons,Song,FD3,0.5 – 1 sec,10,28,38,228,24,0.004
Pitangus sulphuratus,Call,FD3,0.5 – 1 sec,26,47,73,222,26,0.005
Leptotila rufaxila,Song,FD3,0.5 – 1 sec,8,54,62,269,420,0.075
----- Variable duration,,,Repet. interval,,,,,,
Ramphastos tucanus,Song,VD1,0.5 sec,26,68,94,323,44,0.008
Trogon melanurus,Song,VD1,0.7 sec,50,43,93,331,38,0.007
Capito aurovirens,Song,VD1,0.5 sec,283,30,313,449,90,0.016
Glaucidium brasilianum,Song ,VD1,0.3 sec,44,43,87,265,33,0.006
Amazona festiva,Call,VD2,irregular,54,45,99,305,213,0.038
Cacicus cela,Call,VD2,irregular,94,17,111,229,95,0.017
Crotophaga major,Call,VD2,irregular,8,19,27,216,22,0.004
Monasa nigrifrons,Song,VD2,irregular,91,44,135,300,250,0.044
\end{filecontents*}
\csvautotabular {tabs/m01_b_descriptives_full.csv}}
		\caption{Details on training and test data.}
	\end{table}

	\newpage

	\begin{figure}[H]
		\fbox{\includegraphics[width=0.9\textwidth]{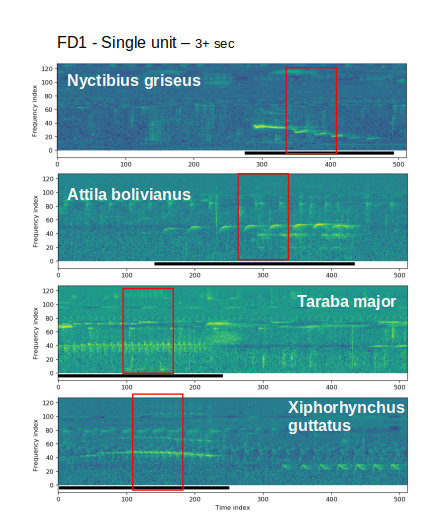}}
		\caption{Spectrograms used as CNN input. Sound-types that were equal or longer than 3 seconds. Time-indexed labels used as target during training shown as black bars at the bottom of each spectrogram. Manually selected Regions of 1.5 seconds shown as red boxes to allow visual comparison with the time scale of acoustic patterns.}
		\label{fig:spe01}
	\end{figure}
	
	\begin{figure}[H]
		\fbox{\includegraphics[width=0.9\textwidth]{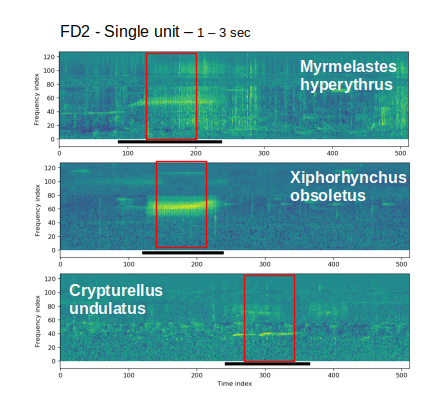}}
		\caption{Spectrograms used as CNN input. Sound-types that were between 1 and 3  seconds.}
		\label{fig:spe02}
	\end{figure}
	
	\begin{figure}[H]
		\fbox{\includegraphics[width=0.9\textwidth]{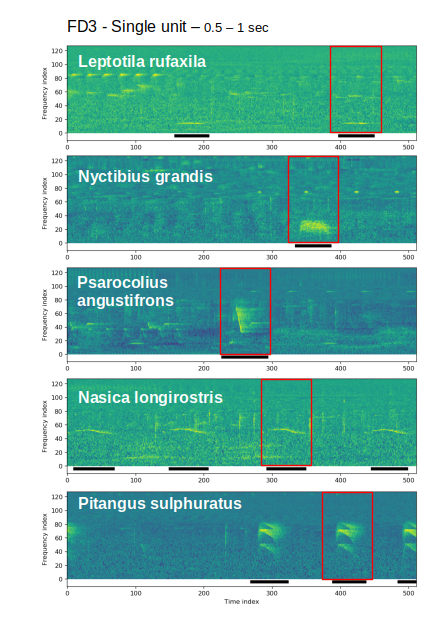}}
		\caption{Spectrograms used as CNN input. Sound-types that were between 0.5 and 1  seconds.}
		\label{fig:spe03}
	\end{figure}	

	\begin{figure}[H]
		\subfloat[\centering AUC]{
			\fbox{{\includegraphics[width=0.5\textwidth]{20220119T005410_timing_tflite_n_trds_1_AUC} }}
		}
		\subfloat[\centering AP]{
			\fbox{{\includegraphics[width=0.5\textwidth]{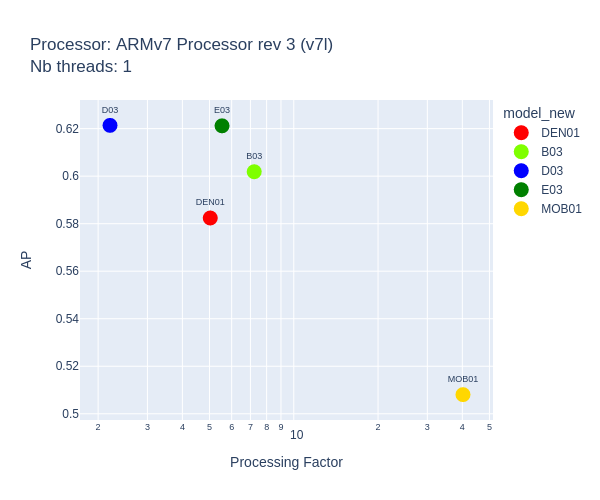} }}
		}
		\caption{Predictive performance of subset of high-performing models plotted against processing factor from inference on a low-performance CPU with a single thread and batch size of 1}%
		\label{fig:timingcpu_appendix}
	\end{figure}

	\begin{figure}[H]
		\subfloat[\centering AUC]{
			\fbox{{\includegraphics[width=0.5\textwidth]{timing_gpu_1_batch_size_32_AUC} }}
		}
		\subfloat[\centering AP]{
			\fbox{{\includegraphics[width=0.5\textwidth]{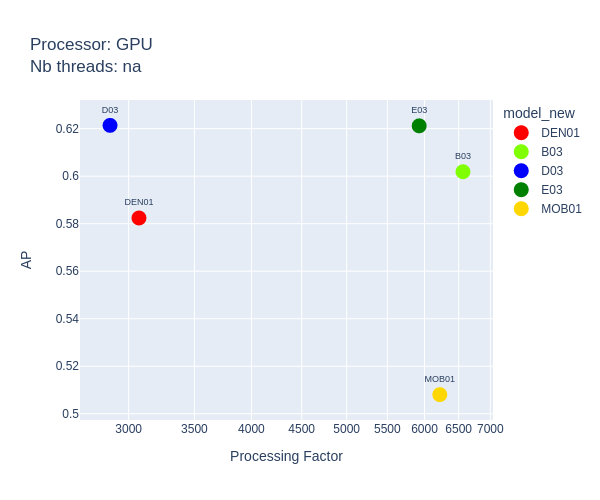} }}
		}
		\caption{Same as Figure 12 but inference on GPU with batch size of 32}%
		\label{fig:timinggpu}
	\end{figure}

\end{document}